\pdfoutput=1
\documentclass[manuscript,screen,nonacm]{acmart}

\setcopyright{none}
\usepackage{amsmath}
\usepackage{quantikz}
\usepackage{tabularx}
\usetikzlibrary{arrows.meta,positioning,calc,fit}

\definecolor{rawblue}{RGB}{0,114,178}
\definecolor{invamber}{RGB}{240,171,0}
\definecolor{calorange}{RGB}{213,94,0}
\definecolor{keyboxbg}{RGB}{245,245,245}
\definecolor{keyboxborder}{RGB}{140,140,140}

\title[Theory-Informed QEM for Shor Order Finding]{Theory-Informed Quantum Error Mitigation for Shor Order Finding with Fourier Neural Operators}

\author{Stefano Markidis}
\author{Qingxin Yang}
\author{Erik {\AA}sgrim}
\author{Ivy Peng}
\affiliation{%
  \department{Department of Computational Science and Technology}
  \institution{KTH Royal Institute of Technology}
  \city{Stockholm}
  \country{Sweden}
}
\renewcommand{\shortauthors}{Markidis et al.}

\begin{document}

\begin{abstract}
Quantum noise distorts the distribution of circuit measurement outcomes, often leaving enough information to infer the algorithm's correct result. To recover this latent information from the noisy distribution in Shor order finding, we introduce a theory-informed quantum error mitigation method. We model how circuit errors shift, broaden, or weaken peaks in the output distribution. These
models guide both the generation of synthetic training data and the
design of a Fourier neural operator with a learned gain that
attenuates or amplifies higher retained Fourier modes. We score each
candidate order by averaging the correlations between its ideal
distribution and the distributions measured in three circuit
executions. The network adjusts these scores to guide the order
estimate and determine the weights of a normalized mixture of the
ideal distributions. On 1,351 simulated test cases for general modular
exponentiation, held out by circuit configuration, exact order recovery
rises from $19\%$ to $34\%$, and the median Jensen-Shannon divergence
from the ideal distribution falls by $50\%$, from $0.64$ to $0.32$.
On 101 Mersenne rotation circuits executed across five IBM quantum
processors, a separate model trained and calibrated in simulation
increases the number of correct order estimates from 35 to 53.
Algorithmic structure can remain recoverable even when noise makes
it barely visible.
\end{abstract}


\maketitle

\section{Introduction}
Reliable quantum computation requires preserving the superposition, interference patterns, and
entanglement used by the algorithm~\citep{markidis2024quantum} despite physical noise and imperfect
control~\citep{Shor1995,GoogleQuantumAI2025}. Quantum error
correction~(QEC) detects and corrects errors during computation. It
encodes each logical qubit, a qubit of the computation, across multiple
physical qubits. Syndrome measurements test whether joint properties of
the physical qubits satisfy the constraints specified by the quantum error code without revealing the logical state. The classical outcomes of these measurements form the error syndrome~\citep{Shor1995,Gottesman1997,Dennis2002}. A decoder is a
classical algorithm that uses the syndrome, the code structure, and a
probabilistic noise model to infer a likely error and select a
correction~\citep{PoulinChung2008}. Quantum error mitigation~(QEM) takes
a different approach. It uses measurements from noisy circuit
executions to estimate quantities of the ideal circuit, such as
expectation values of observables~\citep{Temme2017}. In this work, we
develop a QEM method for reconstructing the ideal output probability distribution from noisy measurements.

Conventional QEC and many QEM techniques describe noise in terms of errors affecting physical qubits, gates, or measurements, such as bit flips, phase flips, gate errors, and readout
errors~\citep{Temme2017,Endo2018}. At the algorithm level, we use known properties of the ideal output to constrain what we infer from noisy measurements. This builds on the use of symmetries and other constraints in error mitigation~\citep{BonetMonroig2018}. The structure of quantum phase estimation has also motivated mitigation methods tailored to that algorithm~\citep{OBrien2021,LiuCai2025}. For instance, in Shor order finding~\citep{Shor1997}, quantum phase estimation produces an ideal measurement distribution with a periodic comb of peaks whose spacing
encodes the order. We model how these errors change the measurement distribution and use the ideal distributions predicted for candidate orders to estimate the order and reconstruct the output distribution. We call this approach \emph{theory-informed algorithm-level QEM}.

Our approach is akin to forward error analysis in classical numerical
computation, which traces how perturbations propagate through an
algorithm and affect its output~\citep{Wilkinson1963,Higham2002}.
We apply this idea to quantum circuits by modeling how errors during
execution change the distribution of measurement outcomes. The theory
of the algorithm tells us what patterns to expect in the ideal output,
such as regularly spaced peaks or probability concentrated on a subset
of outcomes. The error model describes how noise distorts these
patterns or introduces competing ones.

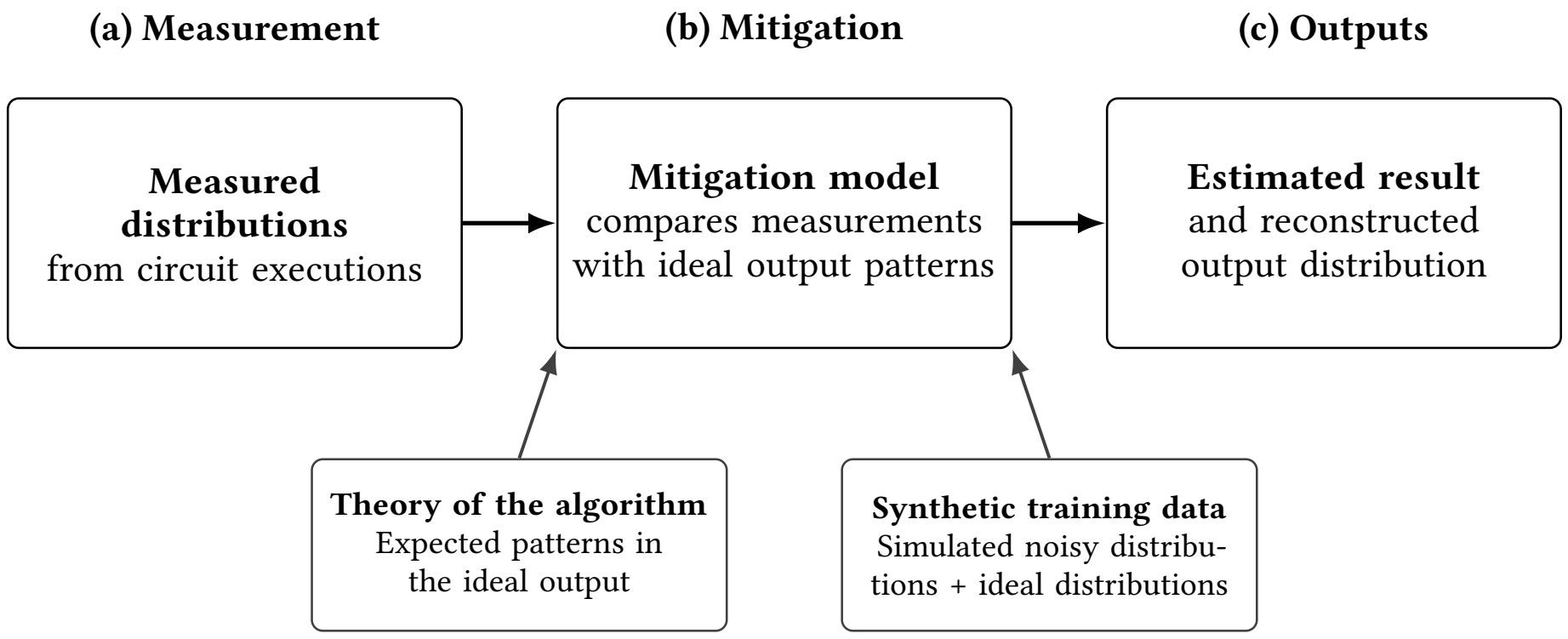
\begin{figure}[t]
\centering
\resizebox{0.8\linewidth}{!}{%
\begin{tikzpicture}[
    node distance=6mm,
    stage/.style={
        draw=black, fill=white, rounded corners=2pt,
        align=center, font=\footnotesize,
        text width=2.70cm, minimum height=16mm,
        inner xsep=3pt, inner ysep=3pt
    },
    input/.style={
        draw=black!75, fill=white, rounded corners=2pt,
        align=center, font=\scriptsize,
        text width=2.45cm, minimum height=11mm,
        inner xsep=3pt, inner ysep=2pt
    },
    stageLabel/.style={font=\footnotesize\bfseries, align=center},
    arrow/.style={-{Latex[length=1.8mm]}, thick, draw=black},
    supportArrow/.style={-{Latex[length=1.6mm]}, semithick, draw=black!75}
]

\node[stage] (measurement)
    {\textbf{Measured}\\
     \textbf{distributions}\\
     from circuit executions};

\node[stage, right=of measurement] (recovery)
    {\textbf{Mitigation model}\\
     compares measurements\\
     with ideal output patterns};

\node[stage, right=of recovery] (decoder)
    {\textbf{Estimated result}\\
     and reconstructed\\
     output distribution};

\node[stageLabel, above=2mm of measurement] {(a) Measurement};
\node[stageLabel, above=2mm of recovery] {(b) Mitigation};
\node[stageLabel, above=2mm of decoder] {(c) Outputs};

\node[input, below=7mm of recovery, xshift=-17mm] (theory)
    {\textbf{Theory of the algorithm}\\
     Expected patterns in\\
     the ideal output};

\node[input, below=7mm of recovery, xshift=17mm] (data)
    {\textbf{Synthetic training data}\\
     Simulated noisy distributions + ideal distributions
     };

\draw[arrow] (measurement) -- (recovery);
\draw[arrow] (recovery) -- (decoder);
\draw[supportArrow] (theory.north) -- (recovery.south west);
\draw[supportArrow] (data.north) -- (recovery.south east);
\end{tikzpicture}%
}
\caption{Overview of theory-informed algorithm-level quantum error
mitigation.}
\Description{Measured circuit distributions enter a mitigation model, which produces an estimated result and a reconstructed distribution. Algorithm theory and synthetic training data support the mitigation model.}
    \label{fig:algorithm_level_mitigation}
\end{figure}
In this work, we introduce a neural network approach for algorithm-level QEM in Shor order finding. The inverse quantum Fourier transform (QFT) in the circuit converts the periodicity of modular exponentiation into a comb of peaks in the ideal output distribution. The spacing between these peaks encodes the order and is reflected in the Fourier spectrum
of the distribution. This motivates our choice of a Fourier neural operator (FNO)~\citep{li2021fourier}, a neural network designed to learn mappings between functions using Fourier layers. The FNO processes measurement distributions from repeated circuit executions. It uses a classical fast Fourier transform (FFT) to compute the Fourier coefficients of the features it extracts from these distributions and applies learned transformations to selected modes. For each candidate order, we compute the ideal output distribution predicted for that order. We assign the candidate a scalar score that measures how closely this distribution matches the measured distributions. The initial score is the average of the correlations with the distributions from individual executions. Higher scores indicate closer agreement. Trained on synthetic data, the FNO predicts how much to increase or decrease each candidate's score. A model parameter sets the maximum
magnitude of this adjustment. We use the revised scores to decide whether to change the initial order estimate. To reconstruct the output distribution, we combine the ideal distributions, giving more weight to candidates with higher revised scores. Figure~\ref{fig:algorithm_level_mitigation} summarizes the mitigation workflow and shows how synthetic training data, analytical reference distributions, and noisy measurements enter the method.

The main contributions of this work are as follows.
\begin{itemize}
\item We introduce an approach to QEM based on models of how errors propagate through a quantum algorithm. For Shor order finding, we model how faults in modular exponentiation, phase errors in the inverse quantum Fourier transform, and readout errors change the distribution of measurement outcomes. These models guide both the generation of synthetic training data and the design of the FNO. 

\item We reconstruct the output as a weighted mixture of the ideal distributions for the candidate orders. Each component retains the peak structure predicted for its order, and the nonnegative weights sum to one, so the reconstruction is a valid probability distribution
by design. On the simulated test set, the median Jensen-Shannon divergence from the ideal output decreases by $50\%$, from $0.64$ for the average of the measured distributions to $0.32$ for the reconstructed distribution.

\item We show that the method improves exact order recovery from $19\%$ to $34\%$ in simulations of general modular exponentiation. On IBM quantum processors, a separate model trained and calibrated entirely in simulation increases the number of correct order estimates
from 35 to 53 out of 101 Mersenne rotation circuits. This transfer from simulation to hardware motivates developing mitigation methods for other quantum algorithms based on how errors change their expected output distributions.
\end{itemize}

\section{Shor Order Finding}
\label{sec:shor-order-finding}
We apply the proposed QEM workflow to quantum order finding, the central quantum subroutine of Shor's factorization algorithm~\citep{Shor1997,EkertJozsa1996}. For an integer $N>1$ and a base $a$ coprime to $N$ (they have no common divisor greater than one), the modular function $a^x\bmod N$ repeats as the exponent $x$ increases. The quantum circuit encodes this repetition coherently and uses interference to transform it into a comb of peaks in the measured probability distribution. The spacing of these peaks reveals the hidden period and therefore provides the algorithm-level structure targeted by our mitigation method.

The period of the modular sequence is called the multiplicative order of $a$ modulo $N$. It is the smallest positive exponent for which the sequence returns to one:
\begin{equation}
r=\operatorname{ord}_{N}(a)
=\min\left\{j\geq1:a^j\equiv1\pmod N\right\}.
\label{eq:order}
\end{equation}
For example, for $N=15$ and $a=2$, the successive modular powers are $1,2,4,8,1,\ldots,$ so the sequence repeats after four steps and the order is $r=4$. The role of the quantum computation is to recover this period from the output distribution.

Once $r$ is known, classical arithmetic can often convert it into factors of $N$. If $r$ is even and $a^{r/2}\not\equiv -1\pmod N$, nontrivial factors are obtained from
\begin{equation}
\gcd\left(a^{r/2}-1,N\right)
\quad\text{and}\quad
\gcd\left(a^{r/2}+1,N\right).
\label{eq:order-to-factors}
\end{equation}
For $N=15$, $a=2$, and \(r=4\), these calculations give $\gcd(3,15)=3$ and $\gcd(5,15)=5$. If the required conditions are not satisfied, the procedure is repeated with another base. Factorization is therefore reduced to recovering the periodicity of modular exponentiation.

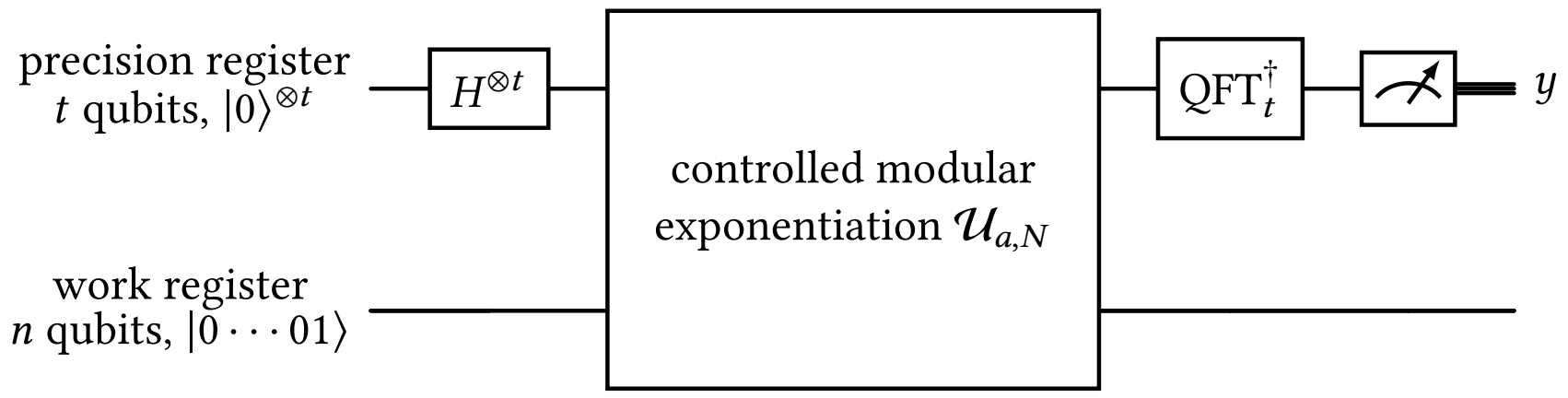
\begin{figure}[t]
\centering

\resizebox{0.75\columnwidth}{!}{%
\begin{quantikz}[
    row sep=0.45cm,
    column sep=0.40cm
]
\lstick{\shortstack{
    precision register\\[-1mm]
    $t$ qubits, $\lvert 0\rangle^{\otimes t}$
}}
&
\gate{H^{\otimes t}}
&
\gate[wires=2]{
    \begin{array}{c}
    \text{controlled modular}\\
    \text{exponentiation } \mathcal U_{a,N}
    \end{array}
}
&
\gate{\mathrm{QFT}^{\dagger}_{t}}
&
\meter{}
&
\cw \rstick{$y$}
\\
\lstick{\shortstack{
    work register\\[-1mm]
    $n$ qubits, $\lvert 0\cdots 01\rangle$
}}
&
\qw
&
&
\qw
&
\qw
&
\qw
\end{quantikz}%
}

\caption{Shor order finding quantum circuit. }
\Description{A precision register passes through Hadamard gates, controlled modular exponentiation, the inverse quantum Fourier transform, and measurement. A work register participates in modular exponentiation.}
    \label{fig:shor-circuit}
\end{figure}

No polynomial-time classical algorithm is known for order finding for general large integers. Shor's quantum procedure requires a number of operations polynomial in the input size~\citep{Shor1997,Lenstra1993}. Figure~\ref{fig:shor-circuit} shows the register-level organization of the quantum subroutine. The $t$-qubit precision register represents $Q=2^t$ possible measurement outcomes, while the $n$-qubit work register, with $n=\lceil\log_2 N\rceil$, stores values modulo $N$. Hadamard gates prepare a uniform superposition of exponents $x=0,\ldots,Q-1$. Controlled modular exponentiation $\mathcal U_{a,N}$ then performs
\begin{equation}
\lvert x\rangle\lvert 1\rangle
\longmapsto
\lvert x\rangle\lvert a^x\bmod N\rangle,
\label{eq:modular-exponentiation}
\end{equation}
thereby encoding the modular periodicity across the two registers~\citep{Vedral1996,Cleve1998,Beauregard2003}. The inverse QFT then converts this encoded periodicity into the characteristic peak comb observed in the precision register. 

The circuit converts the periodicity of modular exponentiation into a measurable peak structure. For an order $r$, the ideal precision-register distribution $p_r(y)$ is concentrated near

\begin{equation}
y_{\ell}\simeq\frac{Q\ell}{r},
\qquad
\ell=0,\ldots,r-1.
\label{eq:order-peak-locations}
\end{equation}
The peak spacing $Q/r$ determines the order $r$. If $r$ divides $Q$, each peak coincides with an integer measurement outcome. Otherwise, some peak locations fall between integers, and their probability spreads across neighboring outcomes even in the ideal circuit. 

A measured outcome $y$ provides an approximation to a fraction whose denominator is related to the order. Shor's classical post-processing uses continued fractions (a representation of numbers using nested reciprocals) to approximate $y/Q$ by a fraction with candidate denominator $d$. It then tests
candidate orders using modular exponentiation~\citep{Shor1997}. When $\ell$ and $r$ share a common factor, the recovered denominator can be a proper divisor of $r$, requiring additional outcomes or tests of denominator multiples.

In our experiments, both decoders estimate the order by comparing the candidates in $\mathcal R_N=\{1,\ldots,N-1\}$. Each candidate $r$ is represented by its analytical distribution $p_r(y)$, and
$r_\star=\operatorname{ord}_N(a)$ denotes the true order.

Both decoders use measurements from three separate executions of the same circuit. Keeping the three distributions separate lets the Fourier neural operator assess which features recur and how much
they vary across executions because of finite sampling and changes in noise. The \emph{raw decoder} uses the same three distributions for comparison. We normalize the counts from each execution
separately as $q^{(e)}(y)=c^{(e)}(y)/S_e$, where $c^{(e)}(y)$ is the count of outcome $y$ in execution $e$, $S_e=\sum_y c^{(e)}(y)$, and $e=1,2,3$.

For an empirical or analytical distribution $f$, we subtract the uniform level $1/Q$ and normalize the resulting vector. The raw score is the mean correlation with each candidate:
\begin{equation}
\begin{aligned}
    \widetilde f(y)
        &=\frac{f(y)-1/Q}{\lVert f-1/Q\rVert_2},\\
    s_r^{\mathrm{raw}}
        &=\frac13\sum_{e=1}^{3}\sum_{y=0}^{Q-1}
          \widetilde q^{(e)}(y)\widetilde p_r(y).
\end{aligned}
\label{eq:fno_raw_score}
\end{equation}
Here, $\lVert\cdot\rVert_2$ is the Euclidean norm over the $Q$ outcomes. Uniform distributions are assigned zero vectors, and small norms are regularized numerically. Centering removes the uniform background, while normalization makes the score invariant to a common positive rescaling of the deviations from that background. Each execution contributes equally to the mean score. The raw decoder returns
\begin{equation}
    \widehat r^{\mathrm{raw}}
        =\operatorname*{arg\,max}_{r\in\mathcal R_N}s_r^{\mathrm{raw}},
    \label{eq:raw_order_estimate}
\end{equation}
with score ties resolved in favor of the smaller order. The FNO approach, used in this work, corrects these raw scores. 

To study this workflow using circuits compatible with current quantum hardware, we exploit the structure of Mersenne integers. Early experiments demonstrated quantum order finding for small instances, including the factorization of $N=15$~\citep{Vandersypen2001}, while general-purpose modular exponentiation requires substantially deeper arithmetic circuits~\citep{Vedral1996}. In hardware experiments, $N=2^n-1$ and the base is a power of two, $a\equiv2^b\pmod N$. Multiplication by $a$ modulo $N$ then becomes a cyclic rotation of the $n$-qubit work register for basis states $0\leq w<N$:
\begin{equation}
U_{a,N}\lvert w\rangle
=
\lvert\operatorname{rot}_b(w)\rangle,
\qquad
r=\frac{n}{\gcd(n,b)}.
\label{eq:mersenne-rotation}
\end{equation}
Here, $\operatorname{rot}_b$ rotates the binary representation of $w$ cyclically to the left by $b$ positions. This construction preserves the periodic order-finding signal while replacing general modular arithmetic with a shallower rotation-based circuit. 

\section{Synthetic and Hardware Datasets}
\label{sec:distribution_datasets}
\begin{figure}[t]
    \centering
    \includegraphics[width=\linewidth]{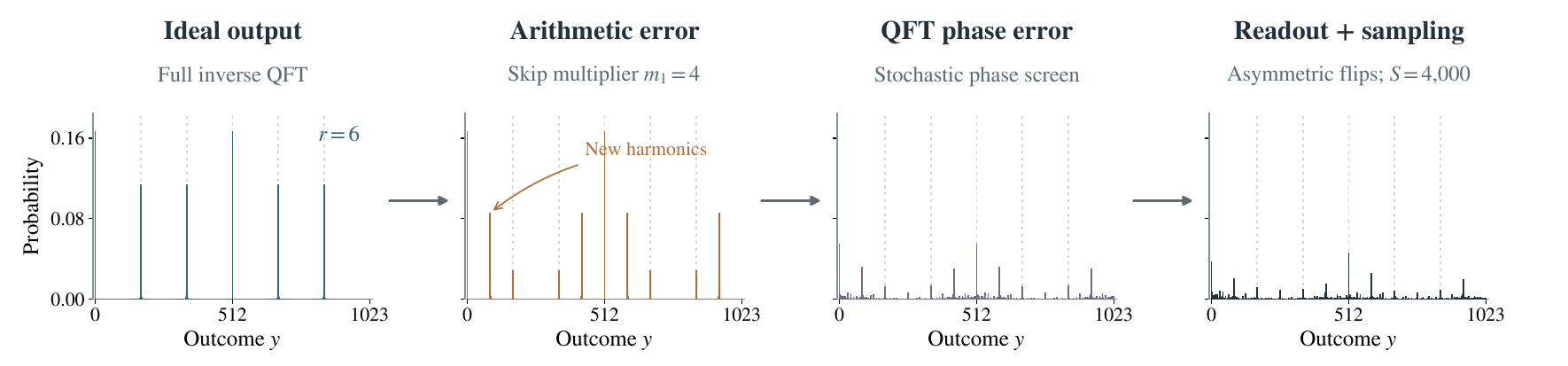}
    \caption{Application-level error propagation for general modular exponentiation with $N=21$, $a=2$, and $Q=1024$. A skipped controlled multiplier creates additional harmonics. Stochastic phase errors, asymmetric readout errors, and 4,000-shot sampling further deform the output. Dashed guides mark the ideal order-six peak family.}
    \Description{Four distributions show the ideal order-six peaks and their successive distortion by an arithmetic fault, stochastic phase errors, and readout errors with finite-shot sampling.}
    \label{fig:error_propagation_spectral_damping}
\end{figure}

In this work, we generate synthetic datasets using statevector simulation with Monte Carlo sampling of errors. Each trajectory represents one sampled sequence of arithmetic and phase errors. During modular exponentiation, the simulator tracks the work-register value and accumulated phase associated with every exponent $x$. It then transforms the resulting amplitudes and traces out the work register to obtain the precision-register probabilities. For the general dataset, the full inverse QFT is evaluated using a FFT. Probabilities are averaged over 16 trajectories before applying readout errors and multinomial sampling. Each example contains three separate 4,000-shot executions and an analytical target computed with the full inverse QFT. Figure~\ref{fig:error_propagation_spectral_damping} illustrates the general error model.

\subsection{Synthetic Data for General Order Finding}
\label{sec:general_modular_dataset}
In ideal modular exponentiation, the work-register label $f(x)=a^x\bmod N$ repeats with the order $r$. Exponent states separated by multiples of $r$ therefore share the same work-register state and retain their coherence when that register is traced out. The inverse QFT maps these repeated separations into a family of peaks near $y=kQ/r$. When $r$ does not divide $Q$, the peaks spread over neighboring integer outcomes. 

\medskip
\noindent \textbf{Modular Exponentiation Error Model.} The modular exponentiation error model perturbs the controlled multipliers $m_j=a^{2^j}\bmod N$. Writing $x=\sum_{j=0}^{t-1}2^jx_j$, a trajectory with multiplier faults alone implements
\begin{equation}
    f_\omega(x)=
    \prod_{j=0}^{t-1}\widetilde m_{\omega,j}^{\,x_j}\bmod N.
    \label{eq:dataset_modular_fault_map}
\end{equation}
In the ideal circuit, block $j$ multiplies the work-register value by $m_j$ modulo $N$ when the control bit $x_j=1$. A skipped block performs no multiplication, corresponding to $\widetilde m_{\omega,j}=1$. An incorrect block uses a different multiplier, chosen coprime to $N$ so that the operation remains reversible. Between blocks, sampled Pauli $X$ errors flip selected work-register bits, while Pauli $Z$ errors change the signs of amplitudes according to those bits. 

One error sequence is sampled for each trajectory and applied to the entire quantum state. Therefore, the random choices are shared across the exponent superposition, rather than sampled independently for each value of $x$. Different trajectories represent different error sequences, whose output probabilities are averaged. Modular fault probabilities increase with work-register width using $p_{\mathrm{eff}}=1-(1-p)^{g_N}$, where $g_N=(\lceil\log_2N\rceil/4)^{2.5}$, before probability caps are applied.

These changes affect which exponent states can interfere after the work register is traced out. For a fault trajectory $\omega$, let $f_\omega(x)$ be the final work-register label and $\theta_\omega(x)$ the accumulated phase. With the full inverse QFT, the probability before readout is
\begin{equation}
\begin{aligned}
    A_\omega(y,w)
    &=\frac{1}{Q}
      \sum_{x:f_\omega(x)=w}
      e^{i\theta_\omega(x)-2\pi ixy/Q},\\
    p_\omega(y)&=\sum_w |A_\omega(y,w)|^2.
\end{aligned}
\label{eq:dataset_fault_propagation}
\end{equation}
Only exponent states ending in the same work-register state interfere in each amplitude. An arithmetic fault can remove separations supporting the intended peak family and introduce other separations with coherent support. The resulting distribution can contain weakened intended peaks and competing harmonic families. 

In Figure~\ref{fig:error_propagation_spectral_damping}, $N=21$, $a=2$, and the true order is six. Skipping the controlled multiplier $m_1=4$ gives $f_\omega(x)=2^{x-2x_1}\bmod21$, which repeats every 12 exponent values. For $Q=1,024$, the orange distribution consequently contains additional peaks near $y=85,427,597,939$. These peaks arise from the altered arithmetic map. 

\medskip
\noindent \textbf{Quantum Fourier Transform Error Model.} The QFT phase-error model modifies the relative phases in Eq.~\eqref{eq:dataset_fault_propagation} while leaving the work-register labels unchanged. It applies a stochastic diagonal phase screen with single-control terms proportional to $x_j$ and pair terms proportional to $x_jx_k$. The coefficients are Gaussian, with pair standard deviations decreasing as $\pi/2^{|j-k|}$. These offsets change constructive interference at the intended harmonics and can attenuate or deform a peak family even when the arithmetic map is correct. The pre-readout distribution $\bar p$ is obtained by averaging 16 coherent trajectories containing the selected arithmetic and phase errors. The third distribution in Figure~\ref{fig:error_propagation_spectral_damping} applies this phase model to the faulty arithmetic map, followed by the full inverse QFT. 

\medskip
\noindent \textbf{Readout Error Model.} The readout model acts on $\bar p$ after the inverse QFT through independent asymmetric bit-flip errors, with separate $0\rightarrow1$ and $1\rightarrow0$ probabilities. It mixes outcome probabilities without changing the preceding arithmetic or interference. This mixing need not be local in the integer outcome index. Flipping bit $j$ moves an outcome by $2^j$. Let $\mathsf A^{(e)}$ be the readout transition matrix for execution $e$, with entry $\mathsf A^{(e)}_{yz}$ giving the probability of recording outcome $y$ given outcome $z$ before readout errors. The
recorded counts follow $\mathbf n^{(e)}\sim\operatorname{Multinomial}(S,\mathsf A^{(e)}\bar{\mathbf p})$,
with $S=4{,}000$ shots per execution. The three executions share $\bar p$ but have perturbed readout rates and independent multinomial samples. The rightmost distribution in Figure~\ref{fig:error_propagation_spectral_damping} shows a normalized count distribution after these final two stages.

The dataset contains 9,399 examples from 497 $(N,a)$ configurations across 26 odd composite moduli in $15\leq N\leq127$. We exclude prime powers because their prime factors can be recovered efficiently by
classical integer root extraction~\citep{Shor1997}. At most 20 coprime bases are selected per modulus, giving 23 distinct orders between 2 and 48. The precision policy $t=2\lceil\log_2N\rceil$ yields $t\in\{8,10,12,14\}$. 

We alternate between $N\leq63$ and $N>63$ to give both ranges comparable representation, and cycle through the available orders within each range to avoid concentrating examples on a few orders.
A noise channel is a component of the simulation that introduces one class of errors. We use three channels to model arithmetic errors in modular exponentiation, phase errors associated with the inverse QFT, and readout errors. Approximately $84\%$ of the examples combine all three channels to capture their joint effect on the output distribution. For each channel, approximately $5\%$ of the examples
activate only that channel to isolate its effect. The remaining $1\%$ have no injected errors and reflect variation from finite sampling alone.

\subsection{Synthetic Data for Order Finding with Mersenne Moduli}
\label{sec:mersenne_modular_dataset}
The Mersenne dataset uses a different arithmetic error model for $N=2^n-1$ and $a=2^b\bmod N$. In this circuit family, modular multiplication is implemented by cyclic rotation of the $n$-qubit work register. The block controlled by precision qubit $j$ rotates by $h_j=b2^j\bmod n$ positions, giving
\begin{equation}
    H(x)=\sum_j h_jx_j=b x\pmod n.
    \label{eq:dataset_mersenne_shift}
\end{equation}
Starting from $\lvert1\rangle$, the work state repeats after $r=n/\gcd(n,b)$ exponent steps.

The arithmetic channel perturbs these rotations through skipped blocks, one-position over- or under-rotations, and erroneous adjacent work-qubit SWAPs. It also includes control-condition toggles, work-register $X/Z$ masks, and control-conditioned phase offsets. These faults replace the general model's incorrect-multiplier mechanism with errors in the rotation implementation. The phase and readout channels retain the forms described above. Every input additionally uses a degree-two approximate inverse QFT, while its clean target uses the full transform.

We use the two-qubit depth and gate count of the corresponding transpiled circuit to adjust how frequently faults are injected in the simulation. When these metrics are unavailable, we base the adjustment on the estimated cost of the logical rotations. Hardware outcome probabilities are not used in this scaling. Each execution independently perturbs the arithmetic and phase parameters, averages 16 trajectories, and applies readout errors and finite-shot sampling. Thus, execution variability also enters the pre-readout distribution in this specialized dataset.

The dataset contains 9,499 examples over 18 $(N,a)$ configurations, with $N\in\{3,7,15,31,63,127\}$, selected bases $a\in\{2,4,8,16\}$, and orders $r\in\{2,3,4,5,6,7\}$. Each configuration is generated at $t=8$ and $t=10$, giving 36 groups sampled cyclically. Approximately 85\% of examples combine the stochastic channels, and approximately 5\% isolate each channel. The degree-two inverse QFT remains present in every category.

\subsection{IBM Quantum Computer Dataset}
\label{sec:ibm_quantum_dataset}
The hardware dataset measures the combined distortions of compiled Mersenne order-finding circuits. It contains 101 circuit configurations and 680 executions on \texttt{ibm\_boston}, \texttt{ibm\_kingston}, \texttt{ibm\_miami}, \texttt{ibm\_pittsburgh}, and \texttt{ibm\_torino}. The moduli are $N\in\{3,7,15,31,63,127\}$, with selected bases $a\in\{2,4,8,16\}$. Of the 101 circuits, 84 use $t=8$ and 17 use $t=10$; the inverse-QFT approximation degree is zero, one, or two. The supplied analytical reference follows the circuit's configured inverse QFT. It is distinguished from the full-QFT reference used by the FNO candidate set and reconstruction.

Each circuit has five or ten separately recorded executions of 4,000 shots. Runs 1-3 provide three-distribution inputs for all 101 circuits. Runs 4-6 and 7-9 provide two further sets of 35 inputs for circuits with ten executions. The later sets repeat existing circuits and retain the variation between hardware runs. Evaluation uses the network weights and score calibration selected on synthetic data, without hardware fine-tuning.

\section{Quantum Error Mitigation with Fourier Neural Operators}
\label{sec:fno_mitigation}
\begin{figure}[t]
    \centering
    \includegraphics[width=\linewidth]{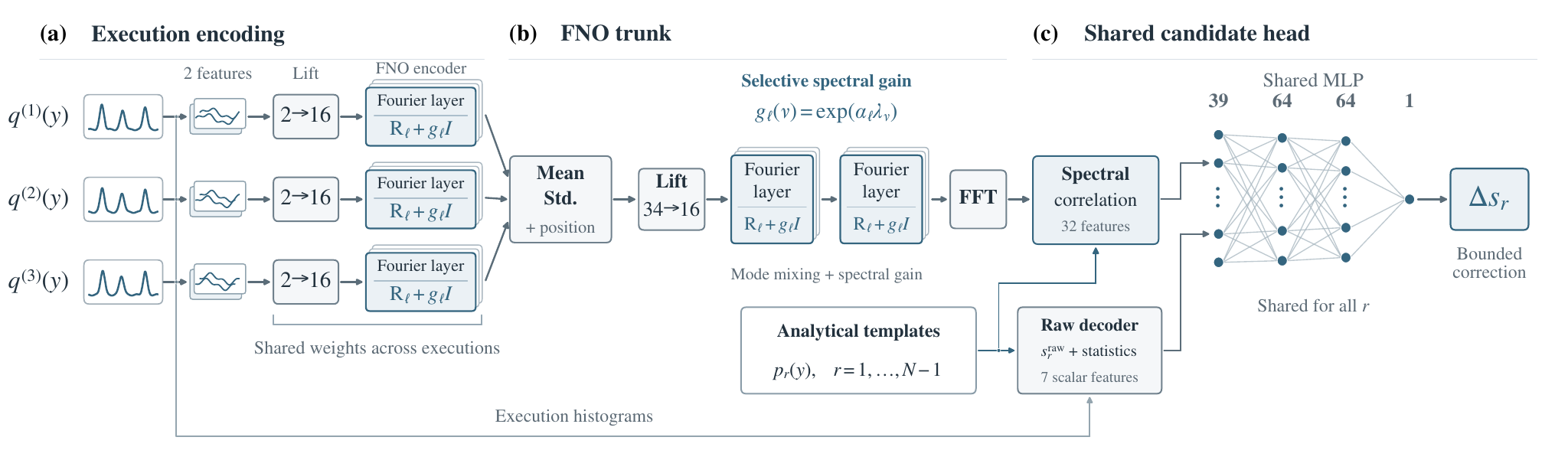}
    \caption{FNO architecture. (a) The same encoder processes each
    measured distribution separately. (b) The trunk processes the
    mean and standard deviation of the resulting features together
    with the outcome coordinates. (c) The candidate head combines
    Fourier comparisons with statistics from the raw decoder to
    adjust the score for each candidate order.}
    \Description{Three measured distributions pass through a shared Fourier encoder. Their feature means and standard deviations enter the trunk with position coordinates. A shared candidate head combines trunk features and raw-decoder statistics to adjust candidate scores.}
    \label{fig:fno_architecture}
\end{figure}

The inverse QFT in Shor order finding produces a pattern of peaks across the measurement outcomes. A classical Fourier transform of the measured distribution represents this pattern through its harmonics. Each Fourier coefficient depends on the distribution across the entire register, so it carries information about the
spacing of peaks even when they are far apart. We use an FNO to extract these features from repeated executions and compare them with the analytical distributions used by the raw decoder. Figure~\ref{fig:fno_architecture} shows how the network turns these comparisons into adjustments to the scores for candidate orders.

\medskip
\noindent\textbf{Combining Information from Repeated Executions.} The encoder is the part of the network that extracts features from one execution. We scale each $q^{(e)}(y)$ by $Q$ so that the uniform level is one at every register size. At each outcome, we use $\log(1+Qq^{(e)}(y))$ to compress large peaks and $\sqrt{Qq^{(e)}(y)}$ to make small nonzero values easier to distinguish from zero. A learned affine map expands these two values into a
larger feature vector. This operation is called a \emph{pointwise lift} because it applies the same map independently at each outcome. A Fourier block then combines information across outcomes.

Using the same encoder for every execution gives corresponding features a common meaning. At each outcome, we compute the mean and standard deviation of each feature across executions, dividing by the number of executions when computing the variance. The mean describes the average pattern. The standard deviation tells the network whether that pattern is consistent across executions or varies substantially. We append $\sin(2\pi y/Q)$ and
$\cos(2\pi y/Q)$ so the network can distinguish positions while respecting the periodic boundary. A learned projection combines these values at each outcome, and two Fourier blocks process the result. These blocks form the \emph{trunk}, the part of the network whose output is computed once and reused for every candidate order.

\medskip
\noindent\textbf{Diffusion and the Design of the Spectral Gain.} Diffusion provides an error model for how peak broadening changes a spectrum. Under diffusion, the decay rate of a Fourier component increases with the square of its frequency, so higher harmonics decay faster. This motivates a gain that depends on the Fourier
mode. Amplification can give more weight to harmonics weakened by broadening. Attenuation can reduce the influence of fluctuations introduced by finite sampling. Each block learns the direction and strength of this adjustment from the training data.

A Fourier block transforms its input features with a FFT~\citep{li2021fourier}. At each retained mode, a learned
complex matrix $\mathsf R_\ell(\nu)$ combines the features, while the gain $g_\ell(\nu)$ multiplies the original coefficients by the same factor for every feature. The block adds these two contributions and transforms them back to the outcome domain. The gain in block
$\ell$ is
\begin{equation}
    g_\ell(\nu)=\exp(\alpha_\ell\lambda_\nu),
    \qquad
    \lambda_\nu=\left(\frac{\nu}{K-1}\right)^2,
    \qquad
    \alpha_\ell=\alpha_{\max}\tanh(a_\ell).
    \label{eq:fno_spectral_gain}
\end{equation}
Here, $\nu=0,\ldots,K-1$ indexes the $K>1$ retained nonnegative modes. The learned scalar $a_\ell$ sets the gain profile. A negative $\alpha_\ell$ gives stronger attenuation at higher modes, and a positive value gives stronger amplification. The parameter $\alpha_{\max}$ keeps the gain between $e^{-\alpha_{\max}}$ and $e^{\alpha_{\max}}$. The encoder applies this operation to each execution, and the trunk applies it to the combined features.

Each block also applies a learned map independently at each outcome. Its output is added to the Fourier result before the nonlinear activation. During training, dropout randomly sets some feature values to zero. The block then adds its input back through a \emph{residual connection} and normalizes across the features at each outcome. Keeping the same number of Fourier modes allows the learned matrices to be reused across register sizes.

\medskip
\noindent\textbf{Adjusting the Scores for Candidate Orders.}
The \emph{candidate head} is the part of the network that produces one score adjustment per order. For each candidate, it compares the Fourier coefficients of each trunk feature with those of the centered and normalized analytical distribution $\widetilde p_r$ from Eq.~\eqref{eq:fno_raw_score}. We compute a normalized complex correlation by multiplying the feature coefficients by the conjugates of the reference coefficients, summing over the selected nonzero modes, and dividing by the product of the two norms over those modes. We regularize small denominators and use the real and imaginary parts separately to retain information about phase alignment.

The head also uses seven quantities from the measurements and circuit dimensions. The mean, standard deviation, minimum, and maximum of the three correlations used by the raw decoder describe how consistently the measurements support the candidate. The gap to the best raw score and the standardized raw score describe its
standing among the candidates. The seventh quantity, $\tanh[\log(Q/N^2)]$, expresses the register resolution relative to the scale used to distinguish rational phases. The correlations from the raw decoder use all outcomes, so they can supply information beyond the selected Fourier modes.

We normalize these inputs and pass them to a multilayer perceptron (MLP), a small network that combines its inputs through successive learned maps and nonlinear activations. The same MLP produces one value $b_r$ for each candidate. We subtract the mean $\overline b$ across candidates to remove a common offset and compute
\begin{equation}
    \Delta s_r=\Delta_{\max}
        \tanh\!\left(\frac{b_r-\overline b}{\Delta_{\max}}\right),
    \qquad
    s_r=\frac{s_r^{\mathrm{raw}}}{T}+\gamma_Q\Delta s_r.
    \label{eq:fno_corrected_score}
\end{equation}
The parameter $\Delta_{\max}>0$ limits the magnitude of the adjustment. The temperature $T>0$ sets the scale of the raw scores while preserving their ranking, and $0\leq\gamma_Q\leq1$ controls how much of the adjustment is applied at each register size. Initializing the final MLP layer to zero makes the network start with the raw ranking.

\medskip
\noindent\textbf{Reconstructing the Distribution and Estimating the Order.} We convert the adjusted scores to weights and use them to combine the analytical distributions:
\begin{equation}
    \pi_r=\frac{\exp(s_r)}{\sum_{j\in\mathcal R_N}\exp(s_j)},
    \qquad
    p^{\mathrm{mit}}(y)=\sum_{r\in\mathcal R_N}\pi_r p_r(y).
    \label{eq:fno_analytical_reconstruction}
\end{equation}
The weights are nonnegative and sum to one, so the reconstruction is a valid probability distribution by design. The analytical components supply the full peak patterns, including details beyond the Fourier modes used by the network.

We use a \emph{switching rule} to decide whether to replace the order with the highest raw score. If a different order has the highest adjusted score, we switch to it only when its adjusted score exceeds the adjusted score of the original order by more than the margin $\eta_Q\geq0$. Otherwise, we keep the original order and the full raw ranking. The reconstruction uses the adjusted scores in either case.

\medskip
\noindent\textbf{Training and Calibration.} The training objective rewards assigning weight to the true order
and matching its analytical distribution, while penalizing large score adjustments:
\begin{equation}
\begin{split}
    \mathcal L={}&
    -\lambda_{\mathrm{ord}}\log\pi_{r_\star}
    +\lambda_{\mathrm H}D_{\mathrm H}^2(p^{\mathrm{mit}},p_{r_\star})
    +\lambda_{\mathrm C}\bigl[
        1-\operatorname{corr}(p^{\mathrm{mit}},p_{r_\star})
    \bigr]\\
    &+\lambda_{\mathrm F}\mathcal L_{\mathrm{spec}}
    +\frac{\lambda_\Delta}{|\mathcal R_N|}
        \sum_{r\in\mathcal R_N}(\Delta s_r)^2.
\end{split}
    \label{eq:fno_training_objective}
\end{equation}
The five $\lambda$ coefficients set the relative weights of the terms. Cross-entropy rewards weight on the true order. Squared Hellinger distance compares the probabilities, including small ones. The centered, normalized correlation compares deviations from uniformity independently of their overall magnitude. The spectral loss compares Fourier coefficients over the evaluated nonzero modes, dividing the squared discrepancy by the sum of the
squared magnitudes of the target coefficients. The final term penalizes the mean squared score adjustment. We train with $\gamma_Q=1$ and apply the switching rule only after training.

\section{Experimental Design}
\label{sec:experimental_design}
This section presents the data partitions, FNO architecture, training and calibration procedures, and evaluation metrics.

\medskip
\noindent\textbf{Data Partitioning.}
We split the synthetic data by circuit configuration $(Q,N,a)$, keeping all noisy realizations of each configuration together. For each $Q$, we shuffle the configurations and assign approximately
$70\%$, $15\%$, and $15\%$ to training, validation, and testing, respectively. In the general dataset, these partitions contain 6,622, 1,426, and 1,351 examples from 348, 75, and 74 configurations,
respectively. The test set contains 18, 15, 627, and 691 examples at $Q=256,1,024,4,096,16,384$, respectively. Each example contributes equally to the reported test rates. We apply the same splitting procedure to the synthetic Mersenne dataset.

The synthetic test set measures generalization to held-out circuit configurations under the same error models used for training. A modulus can appear in more than one partition. The IBM experiments assess transfer to hardware within the Mersenne rotation circuit family. We train and calibrate a separate model on synthetic Mersenne
data, then keep its weights and score calibration fixed during hardware
evaluation. Generalization to entirely held-out moduli and controlled
tests of noise conditions outside the simulation model require
separate evaluation.

\medskip
\noindent\textbf{FNO Architecture.}
We train separate models for the general and Mersenne datasets. Each model has 57,538 trainable parameters and uses the same weights at every register resolution within its dataset. At each outcome, the
encoder maps the two input features to 16 features, and the projection before the trunk maps the 34 values from the aggregated features and position coordinates to 16 features. Each Fourier block retains modes $\nu=0,\ldots,31$ and uses $\alpha_{\max}=1$ to bound the spectral gain. After adding the Fourier and pointwise contributions, we apply a Gaussian error linear unit (GELU). This activation lets the block model
nonlinear relationships between features. It smoothly attenuates negative values and leaves large positive values nearly unchanged.

For each candidate order, the head computes one complex correlation per trunk feature over modes $\nu=1,\ldots,32$. The real and imaginary parts of the 16 correlations give 32 values. We combine them with the
seven quantities defined previously, apply layer normalization, and pass the resulting 39 values to the shared MLP of dimensions $39\rightarrow64\rightarrow64\rightarrow1$. Each hidden layer is followed by GELU and dropout. We initialize the final layer to zero so that training starts from the raw ranking. The spectral term in the
reconstruction loss uses up to 128 nonzero Fourier modes to assess finer detail in the reconstructed distribution.

\medskip
\noindent\textbf{Training and Model Selection.}
We minimize the objective in Eq.~\eqref{eq:fno_training_objective}
using AdamW (an adaptive gradient optimizer that applies weight decay
separately from the gradient update), with batches of eight examples
at the same register resolution. Each $(Q,r_\star)$ stratum receives equal total loss weight, and circuit groups within a stratum receive equal weight. A plateau scheduler halves the learning rate with a patience of two validation checks and a minimum rate of $10^{-6}$. Within each trial, the checkpoint with the highest uncalibrated validation exact-order count is retained, with mean reciprocal rank (MRR), defined below, breaking ties.

Optuna~\citep{akiba2019optuna} optimizes hyperparameters on the fixed validation split using a tree-structured Parzen estimator sampler. Separate searches for the general and Mersenne datasets select optimization parameters, dropout, the score temperature, correction bound, and loss weights. The architecture, training seed, and group assignments remain fixed across trials. Table~\ref{tab:selected_training_parameters} reports the selected settings. The implementation uses \texttt{PyTorch}.

After checkpoint selection, each trial is calibrated on validation data. For each $Q$, we search 51 uniformly spaced strengths $\gamma_Q\in[0,1]$ and switching margins $\eta_Q\geq0$ derived from validation score differences. The selected strengths and margins are included in Table~\ref{tab:selected_training_parameters}.

Trial selection prioritizes calibrated validation exact recovery, with MRR and the fraction of spoiled predictions as secondary criteria. The Mersenne search also considers minimum recovery across resolutions. This comparison is separate from checkpoint selection within a training run. Checkpoint weights and calibration are frozen before test evaluation.

\begin{table}[t]
    \centering
    \caption{Selected optimization, loss, and score parameters. Vector entries follow increasing $Q$.}
    \label{tab:selected_training_parameters}
    \small
    \setlength{\tabcolsep}{4pt}
    \begin{tabular}{lcc}
        \hline
        Parameter & General & Mersenne \\
        \hline
        Learning rate & $7.7332\times10^{-4}$ & $1.7400\times10^{-4}$ \\
        Weight decay & $1.8864\times10^{-4}$ & $4.1517\times10^{-5}$ \\
        Dropout & $0.10$ & $0$ \\
        Gradient clipping norm & $0.5$ & $0.5$ \\
        Raw temperature $T$ & $0.11052$ & $0.10016$ \\
        Correction bound $\Delta_{\max}$ & $4$ & $4$ \\
        $\lambda_{\mathrm{ord}}$ & $4$ & $2$ \\
        $\lambda_{\mathrm H}$ & $0.5$ & $0.5$ \\
        $\lambda_{\mathrm C}$ & $0.2$ & $0.2$ \\
        $\lambda_{\mathrm F}$ & $0.2$ & $0.2$ \\
        $\lambda_\Delta$ & $0.0032514$ & $0.0034787$ \\
        $\gamma_Q$
          & $(0.10,0.72,0.94,1.00)$
          & $(0.94,0.94)$ \\
        $\eta_Q$
          & $(0,0.33548,0.21182,0)$
          & $(0.10011,0)$ \\
        \hline
    \end{tabular}
\end{table}

\medskip
\noindent\textbf{Evaluation Metrics and Procedures.}
 Exact order recovery measures how often the
decoder returns the true order. Top-3 recovery measures how often the
true order has rank at most $\min(3,N_i-1)$ for example $i$. Mean
reciprocal rank (MRR), $M^{-1}\sum_i 1/k_i$, summarizes the full ranking.
A true order ranked first contributes $1$, one ranked second contributes
$1/2$, and so on. Here, $k_i$ is the rank of the true order and $M$ is
the number of evaluated examples.

We rank candidates using the scores selected by the decision rule,
including the raw scores when the switching rule retains the raw
ranking. The rank of a candidate is one plus the number of candidates
whose scores exceed its score by more than $10^{-14}$. Tied scores share
a rank, and an exact tie for the winning score is resolved in favor of
the smaller order. We evaluate the order returned directly by each
decoder, before any continued-fraction processing or arithmetic
verification. Rates for each $Q$ use the examples at that resolution,
and overall rates pool all test examples.

To see where mitigation helps or hurts, we compare the two predictions
for each example. A prediction is \emph{retained} if both decoders are
correct, \emph{recovered} if only the FNO is correct, \emph{spoiled} if
only the raw decoder is correct, and \emph{still incorrect} if both
are wrong. Each percentage in the outcome matrix uses all evaluated
examples at the corresponding $Q$ as its denominator.

In plots of candidate weights,
we exponentiate and normalize $s_{i,r}^{\mathrm{raw}}/T$ for the raw
decoder and $s_{i,r}$ for the FNO. 

We measure how close the reconstructed distribution is to the reference
using Jensen--Shannon (JS) divergence. This divergence compares each
distribution with their average and is zero when they agree.
\begin{equation}
\begin{aligned}
    D_{\mathrm{JS}}(p,q)
       &= \frac{1}{2}\sum_y p(y)\log\frac{p(y)}{m(y)}
        + \frac{1}{2}\sum_y q(y)\log\frac{q(y)}{m(y)},
        \qquad m=\frac{p+q}{2}.
\end{aligned}
\label{eq:experimental_js_divergence}
\end{equation}
With natural logarithms, the divergence is measured in nats and lies
between $0$ and $\log 2$. Terms with zero probability contribute zero.
In the JS divergence plots, the blue series (\emph{Raw decoder}) shows
the divergence of the mean of the three measured distributions
$\bar q_i=\frac{1}{3}\sum_{e=1}^{3}q_i^{(e)}$, and the orange series
(\emph{Neural operator}) shows the divergence of the FNO reconstruction
$p_i^{\mathrm{mit}}$, both relative to the same reference distribution.

To assess how much the results
depend on sampling, we repeat the evaluation at total shot counts
$S\in\{500,\allowbreak1000,\allowbreak2000,\allowbreak4000,\allowbreak8000,\allowbreak12000\}$. For each $S$, we draw 20
subsamples without replacement, allocating the shots proportionally
across the three executions and forming a separate distribution for
each execution. The model weights and calibration stay fixed. For
each draw, we compute exact order recovery and median JS divergence
across examples. The curves average these two statistics over the
20 draws. The bands show the empirical 2.5th and 97.5th percentiles
across draws. At $S=12000$, each draw uses all recorded shots and
reproduces the original distributions.

We test how well a short list preserves the
true order and how often the decoder recovers it from that list. We
rank candidates by either their raw correlation scores or a spectral
comparison that emphasizes features shared across executions. For
the spectral comparison, let
$Z_i^{(e)}=\mathcal F[q_i^{(e)}-1/Q_i]$.
At each nonzero mode of the real FFT, we average
$\operatorname{Re}[Z_i^{(e)}(\nu)\overline{Z_i^{(f)}(\nu)}]$
over the three pairs of executions. A mode contributes positively
when its coefficients are aligned in phase across a pair. We compare
the resulting vector with the squared Fourier magnitudes of each
centered analytical distribution,
$\lvert\mathcal F[p_{i,r}-1/Q_i](\nu)\rvert^2$, using cosine
similarity. This calculation uses the measured distributions directly.

For each $L\in\{1,2,4,8,16,32,64\}$, each screening method retains the
$\min(L,N_i-1)$ highest-ranked candidates. True order recall measures
how often the list contains the true order. Restricted exact recovery
measures how often the decoder returns it. To compute restricted
recovery, we take the raw and corrected scores already computed over
the complete candidate set and apply the switching rule within the
list. Error bars show one binomial standard error,
$\sqrt{\widehat p(1-\widehat p)/M}$, with each example counted once.

\section{Results}
\label{sec:results}
This section presents the results on order recovery and distribution reconstruction, followed by analyses of sampling sensitivity, transfer to IBM hardware, and candidate screening. 

\begin{figure}[t]
    \centering
    \includegraphics[width=\textwidth]{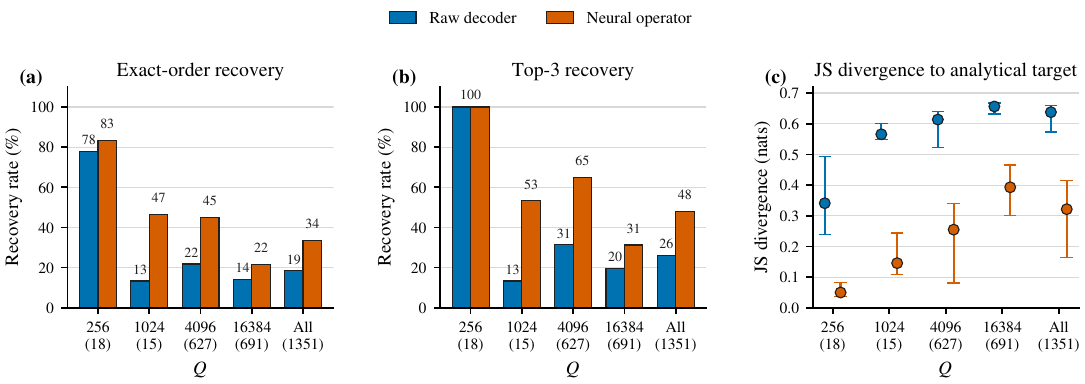}
    \caption{General order-finding results on 1,351 test cases. (a) Exact-order recovery, (b) top-3 recovery, and (c) median JS divergence to the analytical distribution. Error bars indicate the interquartile range.}
    \label{fig:general_results}
    \Description{Three panels compare raw and mitigated exact-order recovery, top-three recovery, and median Jensen-Shannon divergence at each precision-register resolution and over the complete general test set.}
\end{figure}
\begin{figure}[t]
    \centering
    \includegraphics[width=0.8\textwidth]{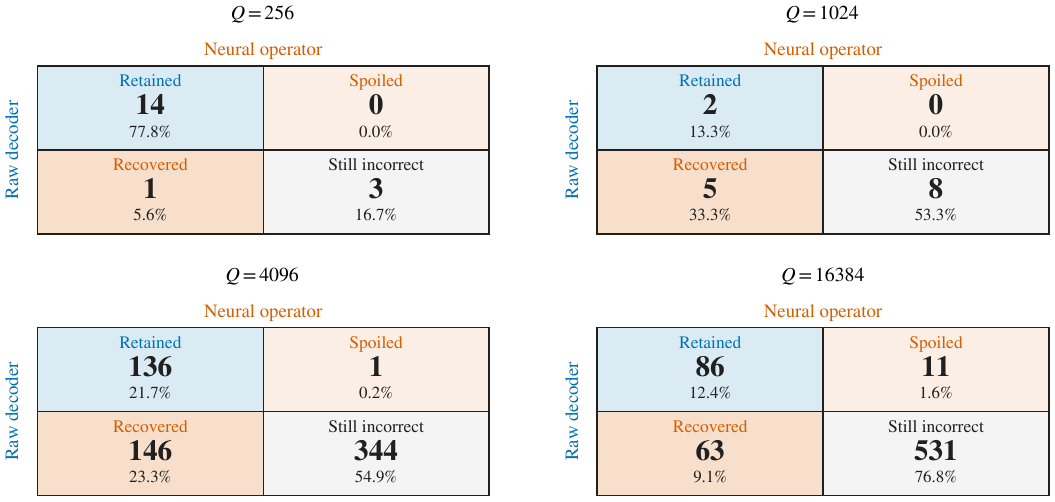}
    \caption{Paired exact-order outcomes on the general test set: retained (both correct), recovered (raw incorrect, FNO correct), spoiled (raw correct, FNO incorrect), and still incorrect (both incorrect). Cells show counts and percentages of test cases at each $Q$.}
    \label{fig:general_order_transitions}
    \Description{Four panels classify raw and neural-operator predictions as retained, spoiled, recovered, or still incorrect, separately for the four register resolutions.}
\end{figure}

\medskip
\noindent\textbf{Overall Order Recovery and Reconstruction.} We first evaluate the effectiveness of FNO-based quantum error mitigation in reconstructing the precision-register distribution and improving order recovery. Figure~\ref{fig:general_results} compares exact-order recovery, top-3 recovery, and JS divergence from the analytical reference distribution across different
register resolutions $Q$. The FNO improves order recovery at all tested resolutions, with the largest gains at $Q=1,024$ and $Q=4,096$. Across all test cases, exact-order recovery increases from 19\% to 34\%, while the median JS divergence decreases from 0.64 to 0.32. However, at $Q=16,384$, exact-order recovery reaches only 22\%, despite the reduction in JS divergence.

To determine how the FNO produces these recovery gains while leaving many cases unresolved, Figure~\ref{fig:general_order_transitions} separates successful corrections from changes that spoil correct predictions. The FNO retains 238 of the 250 correctly decoded raw cases and recovers 215 initially incorrect cases, while spoiling 12, giving a net gain of 203 correct predictions. At $Q=4,096$, it recovers 146 cases and spoils only one. At $Q=16,384$, it recovers 63 cases but spoils 11; both decoders remain incorrect on 531 of the 691 cases (76.8\%). The limited recovery at this resolution therefore reflects predominantly unresolved raw errors, with a smaller contribution from spoiled predictions.

\begin{figure}[t]
    \centering
    \includegraphics[width=0.85\textwidth]{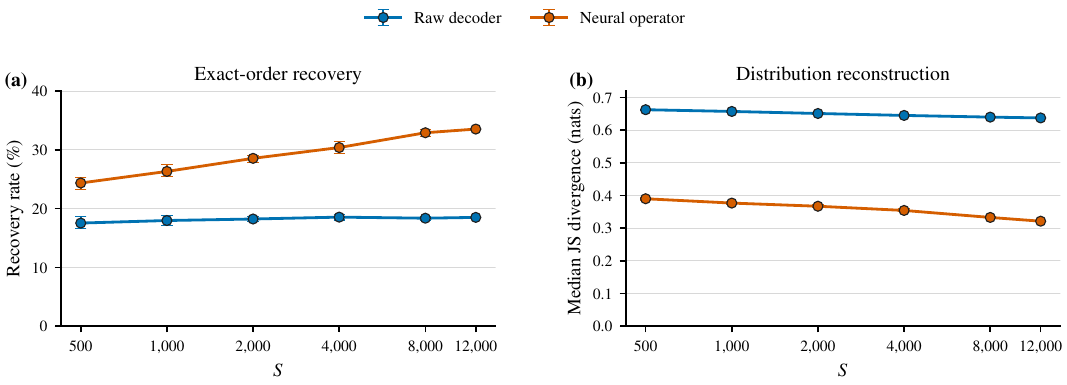}
    \caption{Effect of measurement shots $S$ on
    (a) exact-order recovery and (b) median JS divergence.
    Error bars indicate the 95\% resampling intervals.}
    \label{fig:shot_sensitivity}
    \Description{Two panels show how exact-order recovery and median Jensen-Shannon divergence vary with the number of shots per execution.}
\end{figure}

\medskip
\noindent\textbf{Sensitivity to Measurement Shot Count.} We next investigate the sensitivity of FNO-based mitigation to the number of measurement shots. Figure~\ref{fig:shot_sensitivity} shows that increasing $S$ improves both order recovery and distribution reconstruction. While the raw decoder remains near 18\% exact-order recovery, the FNO increases from 24\%
at 500 shots to 34\% at 12,000 shots. The corresponding decrease in JS divergence indicates that additional measurements allow the FNO to reconstruct the periodic peak structure more accurately. Differently from the raw decoder, the FNO  continues to benefit from increased sampling precision throughout the evaluated range. 

\begin{figure}[t]
    \centering
    \includegraphics[width=0.9\textwidth]{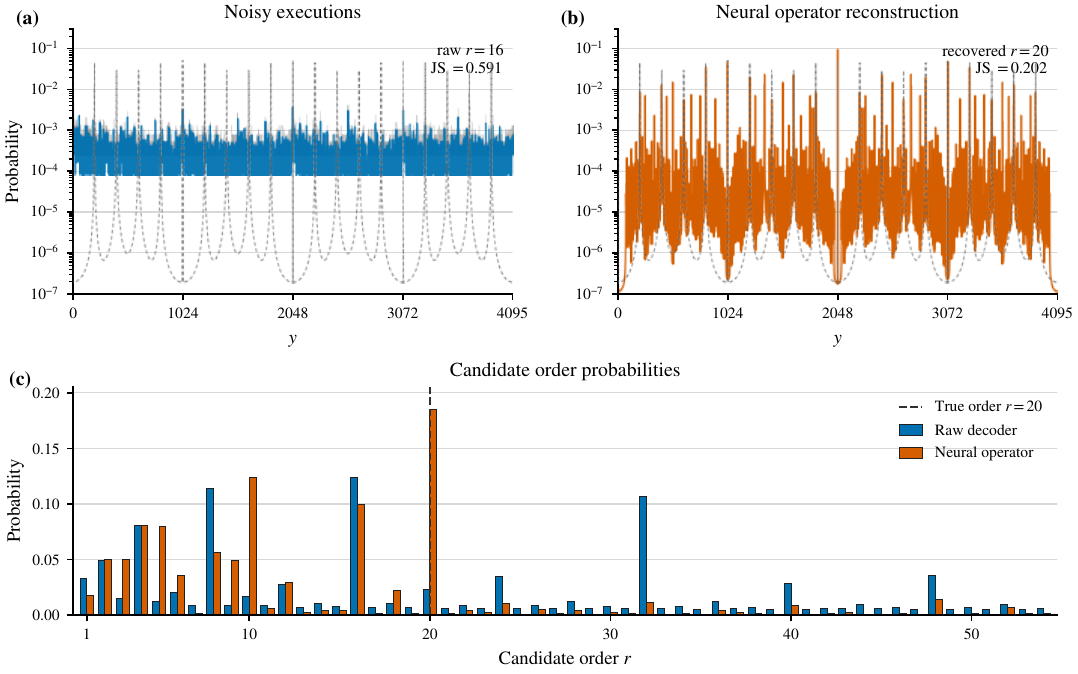}
    \caption{Example of FNO-based reconstruction for $N=55$, $a=3$, $Q=4096$, and true order $r=20$. (a) Three noisy executions (thin gray lines), their mean (blue), and the analytical reference distribution (gray dashed). (b) FNO reconstruction (orange) and analytical reference (gray dashed). (c) Candidate-order probabilities before (blue) and after (orange) mitigation; the vertical dashed line marks the
    true order.}
    \label{fig:reconstruction_example}
    \Description{Three panels compare noisy executions, the reconstructed distribution, and the analytical reference for a representative example with true order twenty.}
\end{figure}

\medskip
\noindent\textbf{Peak Reconstruction and Candidate Ranking.} We next examine how the FNO changes the reconstructed distribution and candidate ranking. Figure~\ref{fig:reconstruction_example} considers an instance with $N=55$, $a=3$, $Q=4,096$, and true order $r=20$. The case is representative of successfully corrected high-resolution instances, with a JS reduction close to the median for this subset. The noisy executions exhibit a nearly uniform probability background that obscures the analytical periodic peak structure. The FNO reconstruction recovers peaks aligned with the analytical reference distribution, reducing JS divergence from $0.591$ to $0.202$. This reconstruction increases the probability assigned to the true order from $2.3\%$ to $18.5\%$, moving $r=20$ from 11th to first in the candidate ranking and replacing the incorrect raw estimate $r=16$. 

\begin{figure}[t]
    \centering
    \includegraphics[width=\textwidth]{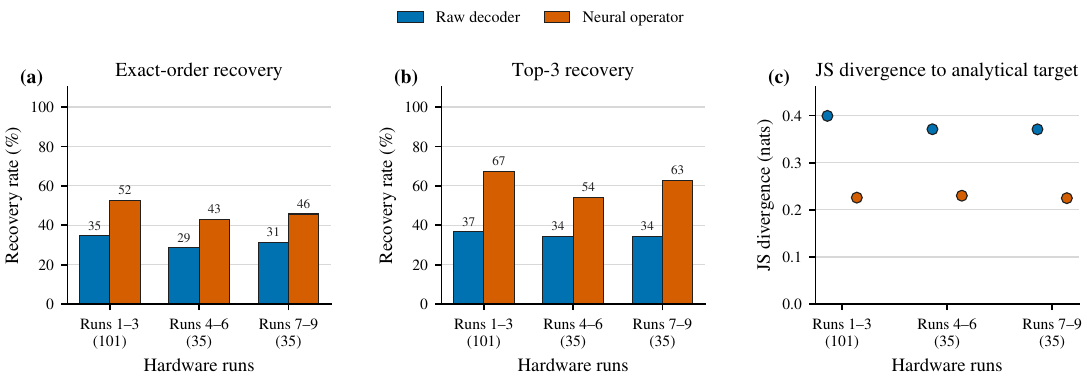}
    \caption{Zero-shot transfer of FNO-based mitigation to IBM quantum hardware. (a) Exact-order recovery, (b) top-3 recovery, and (c) mean JS divergence. Each group contains three independent executions per circuit, covering 101 configurations in runs 1-3 and 35 in each subsequent group.}
    \label{fig:hardware_transfer}
    \Description{Three panels compare exact-order recovery, top-three recovery, and distribution-reconstruction quality across three groups of IBM hardware executions.}
\end{figure}

\medskip
\noindent\textbf{Transfer from Synthetic Data to IBM Hardware.} We next study transfer from simulation to hardware. The FNO is trained on simulated noisy order-finding circuits with Mersenne moduli and applied to hardware measurements without further training. Since the FNO takes three independent measurement distributions as input, we group the hardware executions into sets of three to evaluate the effect of variability between measurements. Figure~\ref{fig:hardware_transfer} compares runs 1-3, covering 101 circuit configurations, with two additional groups covering a subset of 35 configurations. For the complete set, exact-order recovery increases from 35\% to 52\%, top-3 recovery from 37\% to 67\%, and mean JS divergence decreases from 0.40 to 0.23. In both additional groups, the FNO correctly recovers five more orders than the raw decoder, while mean JS divergence decreases to approximately 0.23. 

\begin{figure}[t]
    \centering
    \includegraphics[width=0.9\textwidth]{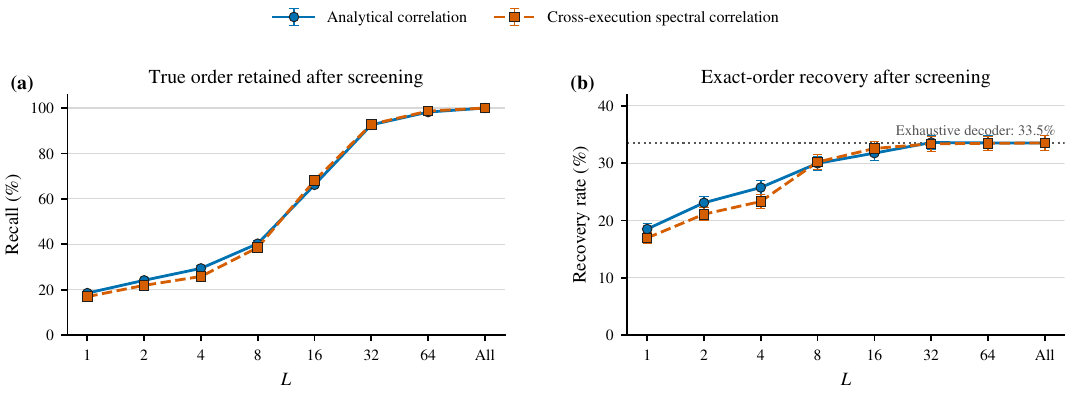}
    \caption{Effect of candidate screening on order recovery. (a) Recall of the true order and (b) exact-order recovery as functions of candidate limit $L$. The dotted line indicates exhaustive decoding. Error bars represent standard errors.}
    \Description{Two panels plot true-order recall and exact-order recovery against the number of retained candidates for analytical-correlation and cross-execution spectral screening. A dotted reference marks exhaustive decoding.}
    \label{fig:candidate_screening}
\end{figure}

\medskip
\noindent\textbf{Spectral Candidate Screening.} Finally, we investigate whether accurate order recovery requires
considering the complete set of candidate orders. This is
important because the number of possible orders grows with
the modulus $N$, whereas the measured periodic structure may
allow the search to be restricted to a smaller set.
Figure~\ref{fig:candidate_screening} evaluates two screening
strategies, based on analytical correlation and spectral
correlation across independent executions, which rank the
candidate orders before retaining the top $L$ for the final
decision. With $L=32$, both methods retain approximately
93\% of the true orders and match the 33.5\% recovery rate
of exhaustive decoding, which considers 71 candidates on
average. 

\section{Discussion}
\label{sec:discussion}
The proposed QEM method could serve as a post-processing layer for circuits that use QEC. Residual logical errors can distort the
distribution of logical measurement outcomes even after error correction. The QEM layer would use this distribution to estimate the ideal output distribution.

\medskip
\noindent\textbf{Selective Candidate Search.}
The order recovery results for general modular exponentiation and Mersenne rotation circuits use exhaustive decoding over $N-1$ candidate orders. The number of candidates therefore grows exponentially with the bit length of $N$. For a fixed network architecture and precomputed analytical distributions, scoring
$R$ candidates and reconstructing the output distribution costs
$\mathcal O(Q\log Q+RQ)$. The Fourier transforms contribute the
$Q\log Q$ term. Comparing the measurements with the analytical
distributions and forming their weighted mixture contribute the
$RQ$ term.

Restricting the decision to the 32 highest-ranked candidates gives
nearly the same recovery rate as exhaustive decoding, even though
the true order is omitted in about $7\%$ of the examples. The
fraction of shortlists that contain the true order sets an upper
bound on recovery, since the decoder can only return an order
that it considers.

In the screening experiment, we compute the raw and adjusted scores over the full candidate set before applying the switching rule within the shortlist. To measure the computational savings of screening, the shortlist would need to be generated before FNO scoring, and its generation cost would need to be included. Scoring only the shortlisted candidates changes the gap to the best raw
score, the standardization of the raw scores, and the mean
subtracted from the MLP outputs. Therefore, recovery under this procedure needs to be evaluated separately.

Continued fractions offer one way to generate candidates directly from the measurements. The denominators of the convergents of $y/Q$ provide candidate orders, and including small multiples accounts for cases where a denominator is a proper divisor of the true order. Seifert proposed simultaneous Diophantine approximation
for order recovery from partial quantum information~\citep{Seifert2001}. Eker{\aa} developed related methods that use limited classical
searches~\citep{Ekera2021Orders,Ekera2024}. For error mitigation, the relevant question is how often these methods include the true order when applied to noisy measurements.

\medskip
\noindent\textbf{Extension to Other Quantum Algorithms.} Extending the method requires a model of how errors change the measurement distribution used to infer the algorithm's result.
Quantum phase estimation is a natural extension because it shares
the inverse QFT and readout stages with Shor order finding.
The corresponding error models can be adapted, while errors in the
controlled operations must be modeled for the circuit at hand.
The analytical references would describe the peaks expected for
candidate eigenphases, with contributions weighted by the input
state's overlap with each eigenspace~\citep{Cleve1998}. The same
construction could be used for amplitude estimation based on phase
estimation~\citep{Brassard2002}.

Quantum walks provide a second example. For a quantum walk, the output distribution gives the probability of
finding the walker at each vertex~\citep{FarhiGutmann1998,Childs2003}.
If noise mainly redistributes probability between neighboring
vertices, graph diffusion provides a model for this effect. Starting
from the ideal distribution, it spreads probability along the edges
of the graph, broadening peaks. Varying the amount of spreading
generates training examples with different levels of distortion.
Diffusion weakens graph Fourier modes that vary sharply between
neighboring vertices. A learned gain could amplify modes weakened by
diffusion and attenuate modes dominated by sampling fluctuations.

\section{Conclusions}
\label{sec:conclusions}
We developed a theory-informed quantum error mitigation method for Shor order finding. Error propagation models guided network design and synthetic data generation. The method improved order recovery and distribution reconstruction on held-out simulated configurations
and on IBM hardware for Mersenne rotation circuits. Further tests should address unseen moduli and noise outside the simulation model.

Future work will extend the hardware dataset to general modular exponentiation and test generalization to larger orders and precision register sizes absent from training. We will assess how often candidate screening retains the true order and how often the decoder recovers it. Experiments with substantially larger precision registers will require sparse representations of the measured distributions. A sparse encoder could compute selected Fourier coefficients directly from these outcomes and their counts. Conditional diffusion~\citep{Ho2020DDPM,Song2021Score} and flow matching~\citep{Lipman2023FlowMatching} could also be explored for reconstruction from noisy measurements. Both could operate on the measured distributions or a reduced Fourier representation and produce a normalized mixture of analytical distributions that preserves the peak patterns predicted for candidate orders.

\begin{acks}
This research was funded by the Wallenberg AI, Autonomous Systems and Software Program (WASP) through the project ``Machine Learning Methods for Reliable Quantum Computing'' and by the Wallenberg Centre for Quantum Technology (WACQT).
\end{acks}

\bibliographystyle{ACM-Reference-Format}
\bibliography{QuantumErrorMitigation}

@inproceedings{Ho2020DDPM,
  author    = {Ho, Jonathan and Jain, Ajay and Abbeel, Pieter},
  title     = {Denoising Diffusion Probabilistic Models},
  booktitle = {Advances in Neural Information Processing Systems},
  volume    = {33},
  pages     = {6840--6851},
  year      = {2020}
}

@inproceedings{akiba2019optuna,
  author = {Akiba, Takuya and Sano, Shotaro and Yanase, Toshihiko and Ohta, Takeru and Koyama, Masanori},
  title = {{Optuna}: A Next-generation Hyperparameter Optimization Framework},
  booktitle = {25th ACM SIGKDD International Conference on Knowledge Discovery and Data Mining},
  year = {2019},
  pages = {2623--2631},
  publisher = {Association for Computing Machinery},
  doi = {10.1145/3292500.3330701}
}

@inproceedings{markidis2024quantum,
  title={What is quantum parallelism, anyhow?},
  author={Markidis, Stefano},
  booktitle={ISC High Performance 2024 Research Paper Proceedings (39th International Conference)},
  pages={1--12},
  year={2024},
  organization={Prometeus GmbH}
}

@inproceedings{li2021fourier,
  title     = {Fourier Neural Operator for Parametric Partial Differential Equations},
  author    = {Li, Zongyi and Kovachki, Nikola and Azizzadenesheli, Kamyar and
               Liu, Burigede and Bhattacharya, Kaushik and Stuart, Andrew and
               Anandkumar, Anima},
  booktitle = {International Conference on Learning Representations},
  year      = {2021},
  url       = {https://openreview.net/forum?id=c8P9NQVtmnO}
}

@inproceedings{Childs2003,
  author    = {Childs, Andrew M. and Cleve, Richard and Deotto, Enrico and Farhi, Edward and Gutmann, Sam and Spielman, Daniel A.},
  title     = {Exponential Algorithmic Speedup by a Quantum Walk},
  booktitle = {Thirty-Fifth Annual ACM Symposium on Theory of Computing},
  pages     = {59--68},
  publisher = {ACM},
  year      = {2003},
  doi       = {10.1145/780542.780552}
}

@inproceedings{Song2021Score,
  author    = {Song, Yang and Sohl-Dickstein, Jascha and Kingma, Diederik P.
               and Kumar, Abhishek and Ermon, Stefano and Poole, Ben},
  title     = {Score-Based Generative Modeling through Stochastic
               Differential Equations},
  booktitle = {International Conference on Learning Representations},
  year      = {2021},
  url       = {https://openreview.net/forum?id=PxTIG12RRHS}
}

@inproceedings{Lipman2023FlowMatching,
  author    = {Lipman, Yaron and Chen, Ricky T. Q. and Ben-Hamu, Heli
               and Nickel, Maximilian and Le, Matt},
  title     = {Flow Matching for Generative Modeling},
  booktitle = {International Conference on Learning Representations},
  year      = {2023},
  url       = {https://openreview.net/forum?id=PqvMRDCJT9t}
}

@article{Ekera2024,
  author  = {Eker{\aa}, Martin},
  title   = {On the Success Probability of Quantum Order Finding},
  journal = {ACM Transactions on Quantum Computing},
  volume  = {5},
  number  = {2},
  pages   = {1--40},
  year    = {2024},
  doi     = {10.1145/3655026}
}

@inproceedings{Seifert2001,
  author    = {Seifert, Jean-Pierre},
  title     = {Using Fewer Qubits in Shor's Factorization Algorithm via Simultaneous Diophantine Approximation},
  booktitle = {Topics in Cryptology---CT-RSA 2001},
  editor    = {Naccache, David},
  series    = {Lecture Notes in Computer Science},
  volume    = {2020},
  pages     = {319--327},
  publisher = {Springer},
  year      = {2001},
  doi       = {10.1007/3-540-45353-9_24}
}

@article{FarhiGutmann1998,
  author  = {Farhi, Edward and Gutmann, Sam},
  title   = {Quantum Computation and Decision Trees},
  journal = {Physical Review A},
  volume  = {58},
  number  = {2},
  pages   = {915--928},
  year    = {1998},
  doi     = {10.1103/PhysRevA.58.915}
}

@article{Ekera2021Orders,
  author  = {Eker{\aa}, Martin},
  title   = {Quantum Algorithms for Computing General Discrete
             Logarithms and Orders with Tradeoffs},
  journal = {Journal of Mathematical Cryptology},
  volume  = {15},
  number  = {1},
  pages   = {359--407},
  year    = {2021},
  doi     = {10.1515/jmc-2020-0006}
}

@article{EkertJozsa1996,
  author  = {Ekert, Artur and Jozsa, Richard},
  title   = {Quantum Computation and Shor's Factoring Algorithm},
  journal = {Reviews of Modern Physics},
  volume  = {68},
  number  = {3},
  pages   = {733--753},
  year    = {1996},
  doi     = {10.1103/RevModPhys.68.733}
}

@article{Vedral1996,
  author  = {Vedral, Vlatko and Barenco, Adriano and Ekert, Artur},
  title   = {Quantum Networks for Elementary Arithmetic Operations},
  journal = {Physical Review A},
  volume  = {54},
  number  = {1},
  pages   = {147--153},
  year    = {1996},
  doi     = {10.1103/PhysRevA.54.147}
}

@article{Vandersypen2001,
  author  = {Vandersypen, Lieven M. K. and Steffen, Matthias
             and Breyta, Gregory and Yannoni, Costantino S.
             and Sherwood, Mark H. and Chuang, Isaac L.},
  title   = {Experimental Realization of Shor's Quantum Factoring
             Algorithm Using Nuclear Magnetic Resonance},
  journal = {Nature},
  volume  = {414},
  pages   = {883--887},
  year    = {2001},
  doi     = {10.1038/414883a}
}

@book{Wilkinson1963,
  author    = {Wilkinson, James H.},
  title     = {Rounding Errors in Algebraic Processes},
  publisher = {Prentice-Hall},
  address   = {Englewood Cliffs, NJ},
  year      = {1963}
}

@article{Cleve1998,
  author  = {Cleve, Richard and Ekert, Artur and Macchiavello, Chiara
             and Mosca, Michele},
  title   = {Quantum Algorithms Revisited},
  journal = {Proceedings of the Royal Society of London. Series A:
             Mathematical, Physical and Engineering Sciences},
  volume  = {454},
  number  = {1969},
  pages   = {339--354},
  year    = {1998},
  doi     = {10.1098/rspa.1998.0164}
}

@article{Beauregard2003,
  author  = {Beauregard, St{\'e}phane},
  title   = {Circuit for Shor's Algorithm Using \(2n+3\) Qubits},
  journal = {Quantum Information and Computation},
  volume  = {3},
  number  = {2},
  pages   = {175--185},
  year    = {2003},
  doi     = {10.26421/QIC3.2-8}
}

@article{Endo2018,
  author  = {Endo, Suguru and Benjamin, Simon C. and Li, Ying},
  title   = {Practical Quantum Error Mitigation for Near-Future Applications},
  journal = {Physical Review X},
  volume  = {8},
  number  = {3},
  pages   = {031027},
  year    = {2018},
  doi     = {10.1103/PhysRevX.8.031027}
}

@article{BonetMonroig2018,
  author  = {Bonet-Monroig, X. and Sagastizabal, R. and Singh, M.
             and O'Brien, T. E.},
  title   = {Low-Cost Error Mitigation by Symmetry Verification},
  journal = {Physical Review A},
  volume  = {98},
  number  = {6},
  pages   = {062339},
  year    = {2018},
  doi     = {10.1103/PhysRevA.98.062339}
}

@article{OBrien2021,
  author  = {O'Brien, Thomas E. and Polla, Stefano and Rubin, Nicholas C.
             and Huggins, William J. and McArdle, Sam and Boixo, Sergio
             and McClean, Jarrod R. and Babbush, Ryan},
  title   = {Error Mitigation via Verified Phase Estimation},
  journal = {PRX Quantum},
  volume  = {2},
  number  = {2},
  pages   = {020317},
  year    = {2021},
  doi     = {10.1103/PRXQuantum.2.020317}
}

@misc{LiuCai2025,
  author        = {Liu, Kecheng and Cai, Zhenyu},
  title         = {Quantum Error Mitigation for Sampling Algorithms},
  year          = {2025},
  eprint        = {2502.11285},
  archivePrefix = {arXiv},
  primaryClass  = {quant-ph}
}

@article{Shor1995,
  author  = {Shor, Peter W.},
  title   = {Scheme for Reducing Decoherence in Quantum Computer Memory},
  journal = {Physical Review A},
  volume  = {52},
  number  = {4},
  pages   = {R2493--R2496},
  year    = {1995},
  doi     = {10.1103/PhysRevA.52.R2493}
}

@phdthesis{Gottesman1997,
  author  = {Gottesman, Daniel},
  title   = {Stabilizer Codes and Quantum Error Correction},
  school  = {California Institute of Technology},
  year    = {1997},
  doi     = {10.7907/rzr7-dt72},
  eprint  = {quant-ph/9705052},
  archivePrefix = {arXiv}
}

@article{Dennis2002,
  author  = {Dennis, Eric and Kitaev, Alexei and Landahl, Andrew
             and Preskill, John},
  title   = {Topological Quantum Memory},
  journal = {Journal of Mathematical Physics},
  volume  = {43},
  number  = {9},
  pages   = {4452--4505},
  year    = {2002},
  doi     = {10.1063/1.1499754}
}

@article{PoulinChung2008,
  author  = {Poulin, David and Chung, Yeojin},
  title   = {On the Iterative Decoding of Sparse Quantum Codes},
  journal = {Quantum Information and Computation},
  volume  = {8},
  number  = {10},
  pages   = {987--1000},
  year    = {2008},
  eprint  = {0801.1241},
  archivePrefix = {arXiv},
  primaryClass  = {quant-ph}
}

@article{Temme2017,
  author  = {Temme, Kristan and Bravyi, Sergey and Gambetta, Jay M.},
  title   = {Error Mitigation for Short-Depth Quantum Circuits},
  journal = {Physical Review Letters},
  volume  = {119},
  number  = {18},
  pages   = {180509},
  year    = {2017},
  doi     = {10.1103/PhysRevLett.119.180509}
}

@article{GoogleQuantumAI2025,
  author  = {{Google Quantum AI and Collaborators}},
  title   = {Quantum Error Correction below the Surface Code Threshold},
  journal = {Nature},
  volume  = {638},
  pages   = {920--926},
  year    = {2025},
  doi     = {10.1038/s41586-024-08449-y}
}

@incollection{Brassard2002,
  author    = {Brassard, Gilles and H{\o}yer, Peter and Mosca, Michele and
               Tapp, Alain},
  title     = {{Quantum} Amplitude Amplification and Estimation},
  booktitle = {Quantum Computation and Information},
  editor    = {Lomonaco, Jr., S. J. and Brandt, H. E.},
  series    = {Contemporary Mathematics},
  volume    = {305},
  pages     = {53--74},
  publisher = {American Mathematical Society},
  year      = {2002},
  doi       = {10.1090/conm/305/05215}
}

@article{Shor1997,
  author    = {Shor, Peter W.},
  title     = {Polynomial-Time Algorithms for Prime Factorization and
               Discrete Logarithms on a {Quantum} Computer},
  journal   = {SIAM Journal on Computing},
  volume    = {26},
  number    = {5},
  pages     = {1484--1509},
  year      = {1997},
  doi       = {10.1137/S0097539795293172}
}

@book{Higham2002,
  author    = {Higham, Nicholas J.},
  title     = {Accuracy and Stability of Numerical Algorithms},
  edition   = {2nd},
  publisher = {Society for Industrial and Applied Mathematics},
  address   = {Philadelphia, PA},
  year      = {2002},
  doi       = {10.1137/1.9780898718027},
  isbn      = {978-0-89871-521-7}
}

@book{Lenstra1993,
  editor    = {Lenstra, A. K. and Lenstra, Jr., H. W.},
  title     = {The Development of the Number Field Sieve},
  series    = {Lecture Notes in Mathematics},
  volume    = {1554},
  publisher = {Springer},
  address   = {Berlin, Heidelberg},
  year      = {1993},
  doi       = {10.1007/BFb0091534}
}

\end{document}